\documentclass[aps,numerical,graphicx,reprint,pre]{revtex4-1}
\usepackage{amsmath}
\usepackage{graphicx}
\draft 
\date{today}
\begin{document}

\def\Journal#1#2#3#4{{#1 }{\bf #2, }{ #3 }{ (#4)}} 
 
\def\BiJ{ Biophys. J.}                 
\def\Bios{ Biosensors and Bioelectronics} 
\def\FNL{ Fluct. Noise Lett.} 
\def\JPS{J. Paramed Sci.}
\def\JCP{ J. Chem. Phys.} 
\def\JAP{ J. Appl. Phys.} 
\def\JMB{ J. Mol. Biol.} 
\def\JPC{ J. Phys: Condens. Matter}
\def\CMP{ Comm. Math. Phys.} 
\def\LMP{ Lett. Math. Phys.} 
\def\NLE{{ Nature Lett.}} 
\def\NPB{{ Nucl. Phys.} B} 
\def\PLA{{ Phys. Lett.}  A} 
\def\PLB{{ Phys. Lett.}  B} 
\def\PNAS{Proc. Natl. Am. Soc.}
\def\PRL{ Phys. Rev. Lett.} 
\def\PRA{{ Phys. Rev.} A} 
\def\PRE{{ Phys. Rev.} E} 
\def\PRB{{ Phys. Rev.} B} 
\def\PD{{ Physica} D} 
\def\ZPC{{ Z. Phys.} C} 
\def\RMP{ Rev. Mod. Phys.} 
\def\EPJD{{ Eur. Phys. J.} D} 
\def\SAB{ Sens. Act. B} 
\def\MST{Meas. Sci Technol.}
\title{Current-voltage characteristics of seven-helix proteins from a cubic array of amino acids} 

\author{Eleonora Alfinito}
\email{eleonora.alfinito@unisalento.it}
\homepage{http://cmtg1.unile.it/eleonora1.html}
\affiliation{Dipartimento di Ingegneria dell'Innovazione,
Universit\`a del Salento, via Monteroni, I-73100 Lecce, Italy}
\author{Lino Reggiani}
\email{lino.reggiani@unisalento.it}
\affiliation{Dipartimento di Matematica e Fisica, "Ennio de Giorgi",
Universit\`a del Salento, via Monteroni, I-73100 Lecce, Italy}

\date{\today}
\begin{abstract}
The electrical properties of a set of seven-helix transmembrane proteins, whose space arrangement (3d structure) is known, are investigated by using  regular arrays of the amino acids.
These structures, specifically cubes, have  topological features similar to those shown by the chosen proteins. 
The theoretical results show a good agreement between  the predicted  current-voltage  characteristics obtained from a cubic array and those obtained from a detailed 3D structure.
The agreement is confirmed by available experiments on bacteriorhodopsin.
Furthermore, all the analyzed proteins are found to share the same critical behaviour
of the voltage-dependent conductance and of its variance.
In particular,  the cubic arrangement evidences  a short plateau of the excess conductance and its variance at high voltages.
The results of the present investigation show the possibility to predict the I-V characteristics of multiple-protein sample even in the absence of a detailed knowledge of their 3D structure.

\end{abstract}


\maketitle 

\section{Introduction}
The interest in the structure and function  of a given protein is crossing the boundaries of biology, joining to  electronics, applied physics and engineering. 
This relevant example of cross-fertilization is due to the quest of finding new and more efficient materials to produce green and  renewable fuels \cite{King12,Renu14}, to reduce wasting \cite{Dai14}, to design highly specific prostheses for medicine \cite{Ghezzi13,Saeedi11}, to insure human and animal safety \cite{Lee12,Edith15}.  
Proteins  perform \textit{in vivo} activities that, when introduced in electronic devices, should allow astonishing  performances \cite{Kobilka98, Ritter09}. 
In particular, many transmembrane proteins act as highly specialized sensors, with levels of sensitivity and selectivity inconceivable for any artifact. 
The use of proteins in nano-bio-devices is, therefore,  the topic of an emerging branch  of electronics, recently introduced as proteotronics\cite{proteotronics}.
Proteotronics has the final objective of producing user-friendly, versatile devices for the daily life so as point-of-care testings and low invasive devices for medicine. 
As a matter of fact,  the final device is conceived to be small and able to produce a clear response. 
This is allowed by the use of quite small active elements (protein samples), and an appropriate miniaturized electronic interface able to convert a biochemical response into an electrical signal.
Recent experiments have shown how the capture mechanism of a ligand (small molecules, odours, photons, etc.) has a solid conjecture concerning the expected protein electrical responses \cite{Lee12,Hou07,Alfinito09d,Ron10,Jin06}.
The microscopic interpretation is actually based on the knowledge of the detailed tertiary  structure of the single protein, i.e. of the space location of its amino acids.
However, from one hand, such a knowledge is largely incomplete and even in the few favorable cases is still a matter of active research far from a general consensus in the scientific community \cite{PDB}.
From another hand, this knowledge is an essential input for modeling the charge transport of protein-based devices \cite{Hou07,Jin06, Melikyan11,Alfinito09, Alfinito11c,Alfinito15}.
Indeed, the specific protein topology produces an electrical response very peculiar of the given protein \cite{Hou07,Lee12,Jin06, Melikyan11}.
Thus, when this input is not known from a credited source,  an approximate structure template able to reproduce the essential features of the electrical response, can be useful to  model the protein electrical properties.  
In particular, we conjecture that for proteins with similar size and shape (globular, fibrous, transmembrane, etc.) the general trend of the electrical response or, at least, of the current-voltage (I-V) characteristic, can be modeled by using a single template.
\par
The aim of this paper is to test the above conjecture.
To this purpose, we have investigated three proteins belonging to the transmembrane family
whose model of the tertiary structure (hereafter simply indicated as the 3D structure) is known \cite{PDB, Alfinito09a}.
For one of them, the bacteriorhodopsin (bR), the I-V characteristic is  also available from experiments\cite{Jin06,Casuso07}.
For these proteins  the I-V characteristic has been calculated using an  impedance network protein analogue (INPA) \cite{Alfinito11c, proteotronics}.
The results obtained from the 3D structure have been compared with those obtained by using a regular structure, here taken as a cube, and proved to capture 
the main features of the protein electrical properties.
The simple cube is built with a number of nodes as close as possible to the number of amino acids given by the public data bank (PDB) \cite{PDB}.
At present only the response corresponding to the protein 
native state will be analyzed.
The presence of the ligand modifies the protein structure and also its free energy \cite{Alfinito15} and this is beyond the aims of this paper.
\begin{figure}
	\centering
		\includegraphics[width=0.46\textwidth]{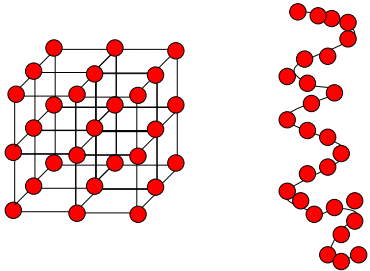}
	\caption{Cartoon (not to scale) of the \textit{cube3 }and of an amino acid helix. Both structures have the same number of nodes. }
	\label{fig:cube1}
\end{figure}
\par
The paper is organized as follows.
In section \ref{sec:model} the theoretical model is briefly recalled.
In section \ref{sec:materials} the given proteins and the analogue cubic structures are introduced.
Section \ref{sec:results} reports the main results of this study and final conclusions are summarized in Sec.~\ref{sec:conclusions}.
\section{Model}
\label{sec:model}
The microscopic approach makes use of the INPA model that describes the single  protein electrical properties starting from the protein topology. 
The model was already detailed in previous papers \cite{Millithaler11, Alfinito12, Alfinito15}
In particular, the amino acid space position, as given by the 3D protein structure,
is used to build up a graph, where each node corresponds to an amino acid. 
Two nodes are then linked if their distance is below a cut-off value, $R_{C}$.
Finally, the links are converted into elemental impedances, with the ensemble of impedances leading to an impedance network specific to describe the electrical properties of the single proteins.
By appropriate scaling, the result of the single protein are compared with available experiments carried out on macroscopic samples.

In the present case, since the objective is the static I-V characteristic, the impedances reduce to resistances.
The observed current response in monolayers of bR validates a tunneling-like mechanism of charge transport, which has been successfully simulated by using a Monte Carlo technique \cite{Alfinito10, Alfinito11c,Alfinito13c}.
The model describes the observed transition between a quasi-linear (Ohmic) regime at low bias, ruled by the direct tunneling mechanism (DT), and a super-Ohmic regime at high bias, mostly due to the injection or Fowler-Nordheim (FN) tunneling  mechanism.
The critical value of the bias controlling the cross-over between the regimes, $V_{C}$, depends on the network structure and corresponds to a microscopic barrier height, $\Phi$, taken to be independent from the specific structure.
The estimated value is $\Phi= 216  meV$ \cite{Alfinito11c}.  

The transition probability for tunneling assumes the standard form: 
\begin{align}
 P^{\rm DT}_{i,j} &=& \exp \left[- \alpha \sqrt{(\Phi-\frac{1}{2}eV_{i,j})} \right] & \label{eq:4a} \\
&&& \rm{if} (eV_{i,j}  < \Phi) , \nonumber \\
{P}^{\rm FN}_{ij}&=&\exp \left[-\alpha\ \frac{\Phi}{eV_{i,j}}\sqrt{\frac{\Phi}{2}} \right] 
  &   
\label{eq:4b} \\
&&& \rm{if} (eV_{i,j} \ge \Phi) \nonumber 
\end{align}
where $V_{i,j}$ is the potential drop between the couple of $i,j$ amino acids, $\alpha =\frac{2l_{i,j}\sqrt{2m}}{\hbar}$, and $m$  is the electron effective mass, here taken the same of the bare value.
When the FN tunneling is applicable $ (eV_{i,j} \ge  \Phi) $, the maximal resistivity decreases, following  the rule:
%
%
\begin{eqnarray}
\rho(V)&=&\rho_{\rm MAX}(\frac{\Phi}{eV_{i,j}})+\rho_{\rm min}(1- \frac{\Phi}{eV_{i,j}}) 
\label{eq:3b}
\end{eqnarray}

where $\rho_{MAX}=4\times 10^{13} \Omega$\AA.\,
 is the  maximum value of resistivity taken to fit the I-V characteristic at the lowest bias, and
$\rho_{min}=4 \times 10^{5}\Omega$\AA\, the minimum resistivity value taken to fit the I-V characteristic at the highest voltages.
\section{Materials}
\label{sec:materials}
We have analyzed three  proteins, namely: bacteriorhodopsin (bR), rat olfactory receptor (OR) I7, and bovine rhodopsin (BR), all in their native state, i.e. the state pertaining to the protein in the absence of a specific external stimulus (ground state) \cite{Alfinito15}. 
As a consequence of an external stimulus,
these proteins may change their native state into an active state.  
Bacteriorhodopsin is analyzed by using its 3D structure, as given by public PDB, in particular the 2NTU entry\cite{PDB}. 
This structure consists of 222 amino acids, exhibits seven helices arranged  to form a loop and has a size of about 5,5 nm. 
\textit{Cube6}, the corresponding regular structure, consists of 6 nodes for edge, making a total of 216 nodes.
The distance between contiguous nodes, the \textit{gen}, is chosen of 5,5 \AA\, which corresponds to a relative maximum of the bR amino acid distance distribution. 
With an analogous procedure the rat OR I7, whose 3D structure  consists of  327 amino acids \cite{Alfinito09a}, is compared with the \textit{cube7}, the corresponding regular structure that consists of 7 nodes for edge, making a total of 343 nodes with \textit{gen}=5,5 \AA .
In both these cases, the value of the interaction radius, $R_C$,  has been chosen of 6 \AA\,  as reported in previous papers \cite{jap, mst}.
The same is done for BR, whose 3D structure consists of  348 amino acids, and the comparison is carried out with the \textit{cube7}.

A cartoon of \textit{cube3} with 3 nodes for edge, and an $\alpha$-helix with the same number of amino acids (27) is shown in Fig.~\ref{fig:cube1}.
\section{Results}
\label{sec:results}
\begin{figure}
	\centering
		\includegraphics*[width=0.45\textwidth]{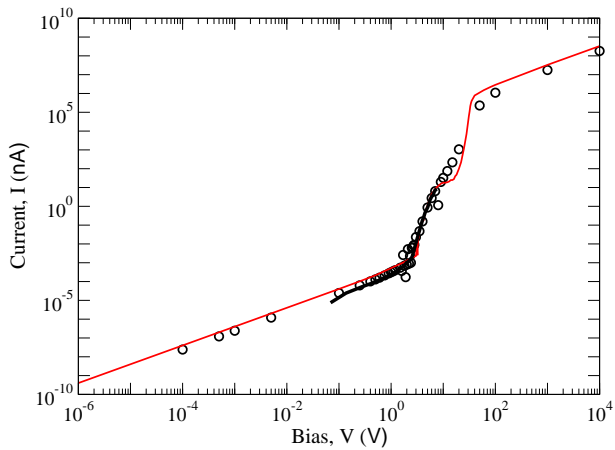}
	\caption{I-V characteristics of bacteriorhodopsin (empty circles) and \textit{cube6}  (continuous red line), $R_C$=6\AA. The bold black line indicates the experimental data \cite{Casuso07}, color online.}
	\label{fig:2ntu-cube}
\end{figure}
\begin{figure}
	\centering
		\includegraphics*[width=0.45\textwidth]{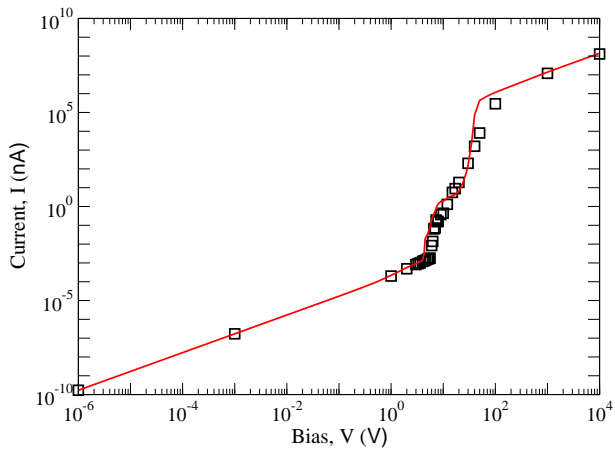}
	\caption{I-V characteristics of rat OR I7 (empty squares) and \textit{cube7} (continuous red line), $R_{C}$ = 6\AA, color online.}
	\label{fig:ORI7_IV}
\end{figure}
\par
The calculated I-V characteristics of \textit{cube6} and bR and of \textit{cube7} and rat OR-I7 are reported in Fig.~\ref{fig:2ntu-cube} and Fig.~\ref{fig:ORI7_IV}, respectively. 

In Fig.~\ref{fig:2ntu-cube} available experimental data  \cite{Casuso07, Alfinito11c} are also shown.
To our knowledge, the measurement of the I-V characteristics on rat OR-I7 samples do not have been performed, yet.
In both cases calculations by using the 3D structure models for these proteins is well reproduced by the corresponding cubes, as well as the DT-FN transition. 
The specific critical bias values are summarized in Tab.~\ref{tab:Table1}. 
\par
The normalized excess conductance,  $\Gamma= \left(\langle g\rangle -g_{0}\right)/g_{max}$, is here taken as the order parameter that describes the associated phase transition between the DT and FN regimes.
With this definition,  $\Gamma$ assumes a value in the interval $(0,1)$.
The normalization is performed by using $g_{0}$, the lowest conductance value (taken at the bias value of 10$^{-6}$ V ) and $g_{max}$, the highest conductance value (taken at the bias value of 10$^{4}$ V) which corresponds to the asymptotic Ohmic behaviour when all the resistivities of the network take the minimum value.

Close to the phase transition, the order parameter calculated for  bR and rat OR I7 exhibits a power-law behaviour, when plotted as a function of the reduced bias, $\epsilon=(V-V_{C})/V_{C}$.
These results are reported in Fig.~\ref{fig:gmaxcube6}.
As is usual in finite-size systems and due to the large fluctuations, rounding and shifting effects \cite{Binder76} do not allow to exactly pinpoint  the phase transition.
Furthermore, since the DT regime overlaps with the FN regime near the transition region, there can be identified two different "critical" bias values, $V_{C}$ and $V_{G}$ \cite{Alfinito13c}. 
The critical bias $V_{C}$ signals the beginning of the FN regime, where the system starts to deviate  from the Ohmic behaviour \cite{Alfinito13c}, while $V_{G}$ , the Ginzburg voltage \cite{Alfinito13c, vitiello},  indicates the consolidation of the FN regime. 

A further signal of the existence of a
universal behaviour is found by analyzing the scaled variance $\left(\langle g^{2}\rangle-\langle g\rangle^{2}\right)/\langle g\rangle^{2}$. 
As just observed by analyzing different proteins \cite{Alfinito13c}, this quantity shows a power law with a critical exponent  close to 3. 
Results are reported in 
Fig.~\ref{fig:variancesacled}.
From the figures above it also emerges  that, better than the current, $\Gamma$  reveals the main differences between the 3D and the cube structure.
Indeed,  while the former shows a continuous increase after the DT-FN transition, the latter exhibits a plateau at intermediate bias. 
We suggest that this plateau is associated with the better
regularity of the cube with respect to the 3D structure. 
As a matter of fact,  by choosing $gen=5,5$ \AA,\, and $R_C$= 6 \AA,\, all the links have the same length and the pattern of potential barriers has itself a cubic shape. 
Conversely, in 3D structures the branches have a wide range of lengths and resistances.
Therefore, as the bias increases (over the FN regime),  at least one link able to get the lowest resistivity is available.
Thus the transition toward the final Ohmic regime takes place continuously.
Conversely, in the chosen cubic structures, the same bias increase produces a stalemate: all the branches have the same resistance value  and none of them is able to prevail and gain the minimal resistivity.
As a consequence, a kind  of critical slowing down is observed at about 3$V_C$.
Finally, this condition is broken and all the  branches achieve the minimal resistivity, and a percolation-like transition  is observed. 
This analysis is represented in figure~\ref{fig:cuberesistivity}, where the scaled resistance, $R/R_{MAX}$, is reported both for \textit{cube6} and \textit{cube7}.
A similar result was obtained by simulating  colossal magnetoresistance in a regular 2D spin lattice, with electrons jumping potential barriers placed in all the sites \cite{Vandewalle}. 
Also in that case, the signature of a percolation transition was identified.
%
\begin{center}
\begin{table}
		\begin{tabular}{l |c|c|c|c|c}
		\hline\hline
		& Cube6 & bR & Cube7 & OR I7 & BR$$\\
			\hline
			\hline
			$V_{C}$& 2,1& 1,7& 2,6& 3,8 & 10\\
			$links$& 540& 692& 882& 1005&1034 \\
			$nodes$ &216&222&343&327& 348\\
		\end{tabular}
	\caption{The critical bias $V_{C}$, links and nodes for the analyzed structures. Voltages are given in Volts, $R_{C}$=6\AA. }
	\label{tab:Table1}
\end{table}
\end{center}
\begin{center}
\begin{table}
		\begin{tabular}{l |c|c|c}
		\hline\hline
		&  Cube7 & OR I7 & BR\\
			\hline
			\hline
			$V_{C}$& 2,6& 1,4 & 2,6\\
			$links$& 882 & 1339 &1417\\
			$nodes$ &327& 343& 348\\
			\hline\hline
		\end{tabular}
	\caption{The critical bias $V_{C}$, links and nodes for the analyzed structures. Voltages are given in Volts, $R_{C}$=7\AA.  }
	\label{tab:Table2}
\end{table}
\end{center}
As a third case of interest, the bovine rhodopsin (BR) has been analyzed. 
This protein 3D structure, entry 1U19 of PDB, consists of 348 amino acids, similar to the rat OR I7.
Therefore, its I-V characteristics should be also described by the \textit{cube7}.
This protein entry gives a structure a little bit more dilated than that of rat OR I7 and
the I-V calculated by using $R_{C}=6$ \AA\ exhibits the value  {$V_{C}= 10$~V,
significantly larger when
compared with the $V_{C}=2,6$~V value obtained with \textit{cube7}, $R_{C}=6$ \AA\, (see Tab.~\ref{tab:Table1}).
On the other hand, by increasing the value of  $R_{C}$ to 7 \AA, both the 3D structure and the \textit{cube7 } give the same value for $V_{C}$, similar to that calculated for rat OR I7 by using  $R_{C}=7$ \AA.
Notice that the results obtained with the cube structure are quite insensitive to the change of the $R_{C}$ value inside the range $(gen, \sqrt2 gen)$.
Results are reported in Fig.~\ref{fig:ORI7-cube_R7} and Table~\ref{tab:Table2}. 
\begin{figure}
	\centering
		\includegraphics*[width=0.45\textwidth]{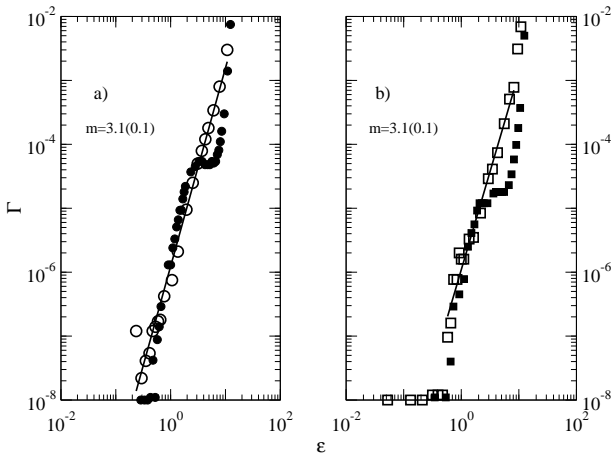}
	\caption{a) Excess conductance for \textit{cube6} (full circles) and bR (open circles), with $R_C$= 6 \AA; b) Excess conductance for \textit{cube7} (full squares) and OR-I7 (open squares), with $R_C$= 6 \AA}
	\label{fig:gmaxcube6}
\end{figure}
\begin{figure}
	\centering
		\includegraphics*[width=0.45\textwidth]{PRE_4.eps}
	\caption{Scaled conductance variance. a) bR, empty circles, and \textit{cube6}, full circles; b) rat OR I7, empty squares, and \textit{cube7}, full squares.}
	\label{fig:variancesacled}
\end{figure}
\begin{figure}
	\centering
		\includegraphics*[width=0.45\textwidth]{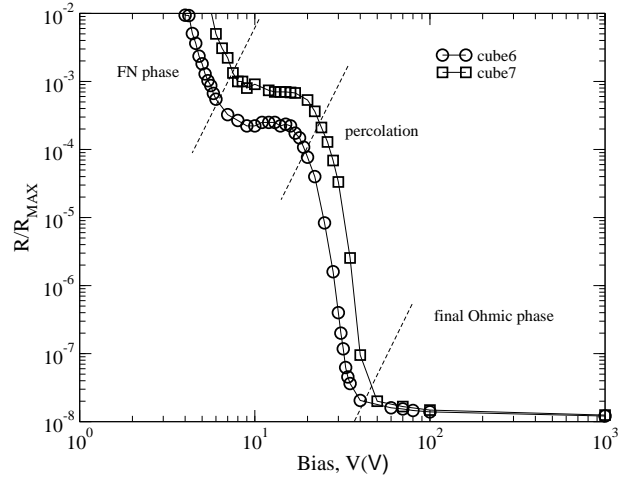}
	\caption{Normalized resistances of \textit{cube6} and \textit{cube7}. }
	\label{fig:cuberesistivity}
\end{figure}
\begin{figure}
	\centering
		\includegraphics*[width=0.45\textwidth]{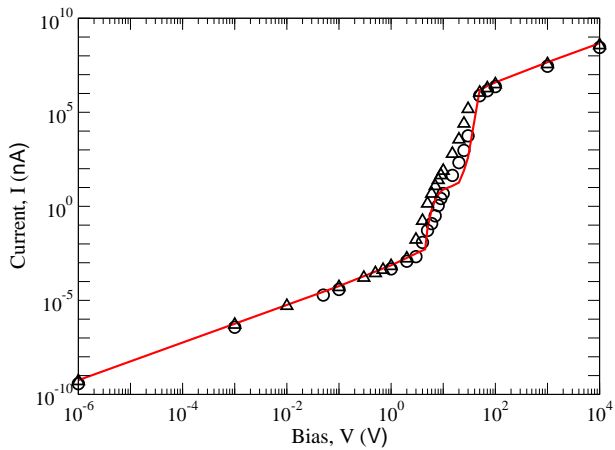}
	\caption{I-V characteristics of BR (circles), rat OR I7 (triangles) and \textit{cube7} (continuous red line), $R_{C}=7$\AA, color online.}
	\label{fig:ORI7-cube_R7}
\end{figure}

In summary, we have found that proteins with a similar number of amino acids exhibit similar I-V characteristics, well reproduced by the cubic structure with \textit{gen=5,5}\AA\,  . 
The choice of a value of $R_{C} = 6 $ or $7$ \AA\,, is irrelevant for the cubic structure, which produces similar responses, but is noteworthy for the 3D structures.
%
\section{Conclusions}
\label{sec:conclusions}
The I-V characteristics of three seven-helix proteins, taken as a prototypes of this family,
 have been investigated within an INPA model.
A comparative analysis between the 3D structure and  a cubic arrangement of 
the amino acids  is then carried out. 
The analysis shows a general agreement between the results of the 3D 
structures and those of the cubic arrays that contain  a similar number of 
amino acids.
In particular, the strong non-linearity of the I-V characteristic is well 
reproduced, together with some interesting peculiarities related to the phase transition 
between a direct and injection tunneling regime of charge transport.
Accordingly,  a cubic protein structure in the form presented here should provide 
realistic expectations of the I-V characteristics ascribed to a given protein even 
in the absence of a detailed knowledge of its 3D structure, thus being  of 
valuable help to predict macroscopic electrical properties.
A further step towards the advancing of proteotronics. 

\begin{acknowledgments}
{This research is supported by the European Commission under the Bioelectronic Olfactory Neuron Device (BOND) project within the grant agreement number 228685-2.}
\end{acknowledgments} 

\end{document}